\documentclass[showpacs,preprintnumbers,amsmath,amssymb,twocolumn]{revtex4}
\usepackage{epsfig}
\usepackage{graphicx}

\begin{document}
\title{Edge states in graphene in magnetic fields \\
--- a speciality of the edge mode embedded in the $n=0$ Landau band}
\author{Mitsuhiro Arikawa$^1$, Yasuhiro Hatsugai$^{1}$ and Hideo Aoki$^2$} 
\affiliation{
$^1$Institute of Physics,  University of Tsukuba, Tsukuba, 305-8571, Japan \\
$^2$Department of Physics, University of Tokyo, Hongo, Tokyo 113-0033, Japan
}

\date{\today}

\begin{abstract}
While usual edge states in the quantum Hall effect(QHE) reside 
between adjacent Landau levels, QHE in graphene has 
a peculiar edge mode at $E=0$ that reside right within 
the $n=0$ Landau level as protected by the chiral symmetry.  
We have theoretically studied the edge states 
to show that the $E=0$ edge 
mode, despite being embedded in the bulk Landau level, 
does give rise to a wave function whose charge is accumulated along 
zigzag edges.  This property, totally outside continuum models, 
implies that the graphene QHE harbors edges distinct from 
ordinary QHE edges with their 
topological origin. 
In the charge accumulation the bulk states 
re-distribute their charge significantly, which may be called a topological 
compensation of charge density.  
The real space behavior obtained 
here should be observable in an STM imaging.   
\end{abstract}

\pacs{68.37.Ef, 73.43.-f}

\maketitle

{\it Introduction ---}  
Ever since the anomalous quantum Hall effect (QHE) was experimentally 
observed,\cite{graphene1,graphene2} 
fascination with graphene is mounting.  
The interests have been focused on the ``massless Dirac" dispersions around Brillouin zone corners (K, K') in 
graphene, where the Dirac cone is topologically protected 
due to the chiral symmetry\cite{topology1}.  
The peculiar dispersion is responsible for the appearance of the 
$n=0$ Landau level ($n$: Landau index) precisely around energy $E=0$ 
in magnetic fields.  For the ordinary integer QHE 
an important general question 
is how the bulk and edge QHE conductions are related for finite samples.  
Many authors have 
addressed this question,\cite{laughlin,halperin} where one of the 
present authors has shown that 
the bulk QHE conductivity, a topological quantity, coincides 
with the edge QHE conductivity, itself another topological quantity.  
This constitutes a typical example of phenomena that, when a bulk system has a topological order\cite{topologicalorder1,topologicalorder2,topologicalorder3,
topologicalorder4} that reflects the geometrical phase of 
the system\cite{phase}, this should be reflected and become visible in the 
edge states in a bounded system.\cite{BEcorresponds1,BEcorresponds2}  
For graphene, two of the present authors and Fukui have shown that 
this ``bulk-edge correspondence" persists in graphene, with both 
an analytic treatment of the topological numbers and numerical results for 
the honeycomb lattice\cite{topology2,dresden}.

Now, in the physics of graphene, it is important to distinguish between 
the properties that arise from the continuum theory (i.e., the 
massless Dirac dispersion that comes from the $k\cdot p$ perturbation 
in the effective-mass formalism) from the properties that can only be 
captured by going back to the honeycomb lattice.  
In ref.
\cite{topology2,dresden}, we have already recognized this 
in a change from the Dirac to fermionic behaviors at 
van Hove singularities of the honeycomb lattice, and in the QHE edge modes 
that depend on whether the edge is zigzag of armchair.  

\begin{figure}
\centerline{\epsfxsize=8.cm\epsfbox{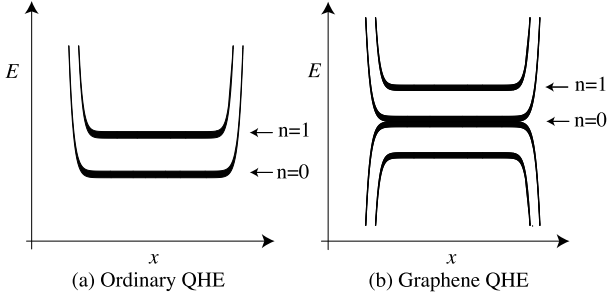}}
\caption{
Schematic Landau-quantized spectra against 
real space position ($x$) for finite systems 
for ordinary QHE systems(a) 
and graphene QHE(b).
\label{schematic}
} 
\end{figure}

The purpose of the present paper is to reveal the features in 
the {\it real-space} profile 
of the edges states in graphene in magnetic fields $B$ 
in the one-body problem.  The graphene edge states in fact turn out to behave 
unusually.  
A crucial point is that we find, from numerically obtained dispersion 
and wave functions, that the $E=0$ edge 
mode, despite being {\it embedded} right within the $n=0$ bulk Landau level 
in the energy spectrum, 
has a wave function whose charge is {\it accumulated} along zigzag edges.  This 
situation is drastically different 
from the ordinary QHE where edge modes 
reside, in energy, between adjacent Landau levels, and 
their charge is depleted toward an edge.  
We can indeed realize in Fig. \ref{schematic} (Landau spectrum vs position) 
that $E=0$ mode in graphene is special.  
The physics here points to a topological origin in a honeycomb lattice, which is in fact 
totally outside continuum models.  
In the absence of magnetic fields, a zigzag edge in graphene 
has been known to have a flat dispersion at $E=0$,\cite{fujita} which 
is protected by the bipartite symmetry of the honeycomb lattice.  
Here we are talking about the edge states in strong magnetic fields, 
which has a flat dispersion at $E=0$.  
The charge accumulation along zigzag edges only occurs 
for the $E=0$ edge mode in the $n=0$ Landau level.  
To be more precise, the bulk states 
re-distribute their charge significantly on top of the zero-mode 
contribution, which may be called a topological 
compensation of charge density.  On a practical side the 
present result predicts how an STM imaging should look like 
for graphene edges\cite{stm}.   
For comparison we have also examined edges in 
a bilayer graphene.\cite{bilayerex1,bilayerex2}

\begin{figure}
\centerline{\epsfxsize=8.5cm\epsfbox{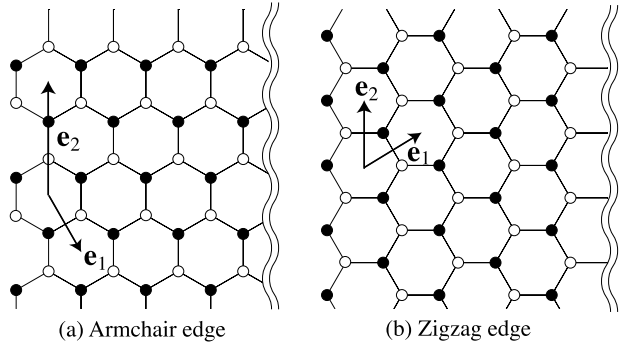}}
\caption{
Honeycomb lattice with armchair (a) or zigzag (b) edges with 
${\mathbf e}_1, {\mathbf e}_2$ being respective unit translation vectors.
\label{labelling}
} 
\end{figure}

{\it Single Layer ---}  
We consider the standard tight-binding model on the honeycomb lattice 
with nearest-neighbor hopping, where 
the magnetic field is introduced as a Peierls phase in the Landau gauge.  
The magnetic field is characterized by the flux 
in units of the magnetic flux quantum,
$
\phi \equiv  B S_6/(2\pi) = 1/q
$
in each hexagon with area $S_6 = (3\sqrt{3}/2) a^2$.
Since we want to look at edges, we should be more 
explicit about the Hamiltonian. 
Since honeycomb is a non-Bravais, bipartite lattice with two sublattice 
sites $\bullet$ and $\circ$ per unit 
cell, we can define two fermion operators 
$c_{\bullet}$ and $c_{\circ}$.  
For an armchair edge (Fig.~\ref{labelling}(a)) 
the Hamiltonian then reads   
\begin{eqnarray*}
{\mathcal H}_A &=t&\sum_{{\mathbf j}} \Big[
c ^\dagger _\bullet({\mathbf j})c_\circ({\mathbf j})
+c ^\dagger _\bullet({\mathbf j}+{\mathbf e}_1)c_\circ({\mathbf j})
\nonumber\\
&&
+e^{i2\pi \phi j_1} c ^\dagger _\circ({\mathbf j}+ {\mathbf e}_1+{\mathbf e}_2)c_\bullet({\mathbf j})
\Big] +\mbox{H.c.}
\end{eqnarray*}
Here ${\mathbf j}= j_1 {\mathbf e}_1+j_2 {\mathbf e}_2$ with 
${\mathbf e}_1, 
{\mathbf e}_2
$ defined in Fig.~\ref{labelling}(a) specifies the position of a unit cell.  
For a zigzag edge (Fig.~\ref{labelling}(b)) 
the Hamiltonian reads
\begin{eqnarray*}
{\mathcal H}_Z &=t&\sum_{{\mathbf j}} \Big[
c ^\dagger _\bullet({\mathbf j})c_\circ({\mathbf j})
+e^{i2\pi \phi j_1}c ^\dagger _\bullet({\mathbf j})c_\circ({\mathbf j}-{\mathbf e}_2)
\nonumber\\
&&
+c ^\dagger _\bullet({\mathbf j}+ {\mathbf e}_1)c_\circ({\mathbf j})
\Big] +\mbox{H.c.}
\end{eqnarray*}
with ${\mathbf e}_1, 
{\mathbf e}_2
$ as defined in Fig.~\ref{labelling}(b).  
Hereafter we take $t$ as a unit of energy and $a$ as a unit of length.

We assume that the system has left and right edges 
with a spacing  $L_1$, taken to be large enough ($L_1= 5q$ here) 
to avoid interference.\cite{footnote1}  
The length along the direction (${\mathbf e}_2$) parallel to the edge 
is also assumed to be long enough ($L_2$), for which we 
apply the periodic boundary condition. 
We can then make a Fourier transform in that direction, 
$
c_{\alpha}({\mathbf j})  =  L_2^{-1/2} \sum_{k_2} e^{i k_2 j_2} c_{\alpha}(j_1,k_2),
$ 
for $j_2 = 1,2,\cdots, L_2$ and $\alpha = \bullet, \circ$. 
This yields a $k_2$-dependent series of one-dimensional Hamiltonian, ${\mathcal H} = \sum_{k_2} {\mathcal H}_{1D}(k_2)$.
 The resultant eigenvalue problem reduces to 
${\mathcal H}_{1D}(k_2) | \psi(k_2,E) \rangle = E | \psi(k_2,E) \rangle $,
 with corresponding eigenstates 
 $|\psi(k_2,E) \rangle$. 

Having STM images in mind, we define the local charge density, 
\begin{eqnarray}
I(x(j_1)) & = & \frac{1}{2\pi}  \int ^{E_2}_{E_1} dE \int
d k_2 
|\psi_\alpha (E,j_1, k_2)|^2. 
\label{defchargedensity}
\end{eqnarray} 
Here $x$ is the distance from the edge 
(as related to $j_1$ via ${\mathbf e}_1$ which is not normal to the edge), 
and $E_1 < E < E_2$ is the energy window to be included in the 
charge density (which is normalized to unity when the window covers the 
whole spectrum).  

\begin{figure}
\centerline{\epsfxsize=7cm\epsfbox{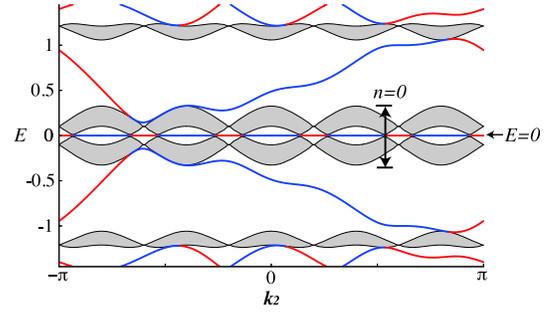}}
\caption{(Color online)
Energy spectra against $k_2$ (momentum along the edge) 
for a single-layer graphene 
in a magnetic field of $\phi=1/5$
with zigzag edges.  
Shaded regions are the bulk energy spectra, while 
red (blue) lines are the modes localized on the zigzag (bearded) edge. 
} 
\label{phi5} 
\end{figure}

\begin{figure}
\centerline{\epsfxsize=8cm\epsfbox{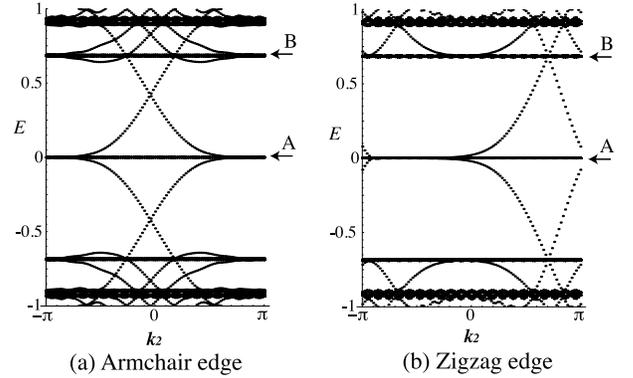}}
\caption{
Energy spectra against $k_2$ (momentum along the edge) 
for a single-layer graphene 
in a magnetic field of $\phi=1/21$
for armchair(a) or zigzag(b) edges.  
} 
\label{fig1} 
\end{figure}

\begin{figure}
\centerline{\epsfxsize=7cm\epsfbox{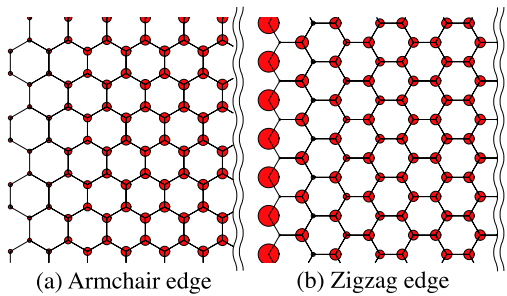}}
\caption{(Color online)
Local charge density ($\propto$ area of each circle) 
for the single layer graphene 
with armchair(a) or zigzag(b) edges 
in a magnetic field $\phi=1/21$ 
for the energy window $-0.05<E<0.05$ around $n=0$ Landau 
level (see Fig.\ref{fig1}).  
} 
\label{fig2} 
\end{figure}

We have stressed that the $E=0$ edge mode is embedded within the 
$n=0$ Landau level, which is depicted in Fig.~\ref{phi5}, 
a blowup of 
the energy spectrum (for a relatively high $\phi=1/5$ for 
clarity).\cite{topology2} 
The shaded regions represent the $n=0$ Landau band, while the red curves 
the edge modes localized along the zigzag edge. 
We can see that, despite the presence of a strong magnetic field, 
there exists an exactly $E=0$ edge mode piercing the 
$n=0$ bulk Landau band. 
We can realize its topological origin by noting that there are an odd 
number ($2q-1$) of edge modes with zig-zag edges, so that the bipartite symmetry %
(that forces an electron-hole symmetric energy spectrum) precisely dictates 
that the central edge mode has to be flat and at $E=0$.\cite{topology2}  

Figure~\ref{fig1} shows the energy spectrum in a magnetic field $\phi=1/21$ 
adopted hereafter for the armchair and zigzag edges.  
For this more moderate field the $n=0$ Landau level around $E=0$, 
with a narrow energy width ($\sim 0.05$), almost looks like a 
line spectrum on this energy resolution.   
We calculate the local charge density defined in 
eq.(\ref{defchargedensity}) for the armchair and zigzag edges 
with the energy window $-0.05<E<0.05$ set to cover the $n=0$ Landau 
level (along with the embedded $E=0$ edge mode).\cite{footnotewidth}  
In the result in Fig.~\ref{fig2} 
the charge density for an armchair edge decreases 
monotonically toward the edge, 
where the depletion occurs 
on the magnetic length scale 
($l_B=3^{3/4}a/\sqrt{2\pi \phi})$), as in ordinary QHE systems.  
In sharp contrast, a zigzag edge has 
the charge density for the $\bullet$-sublattice that is 
{\it accumulated} toward the edge while the charge density for the 
$\circ$-sublattice is depleted.  

\begin{figure}
\centerline{\epsfxsize=8cm\epsfbox{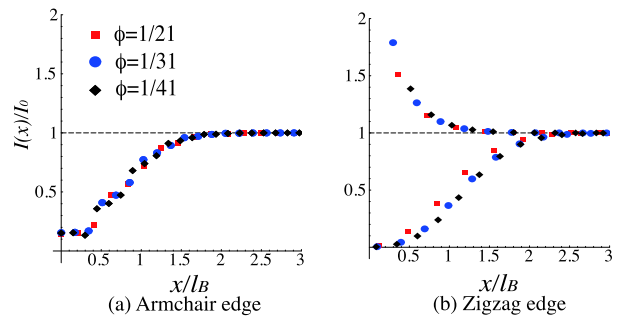}}
\caption{(Color online)
Scaled plot of the charge density $I(x)$ against $x/l_B$, the 
distance from the edge normalized by the magnetic length, 
for the $n=0$ Landau level (marked with A in Fig.~\ref{fig1}) with various values of magnetic field 
$\phi$ for armchair(a) or zigzag(b) edges. 
} 
\label{fig3} 
\end{figure}

The question then is how the 
accumulation of the charge around the zigzag edge 
scale with the magnetic field.   
We plot in Fig.\ref{fig3} the charge density $I(x)$ normalized by 
the bulk value $I_0$ (which is $\phi$, when 
each Landau level of massless Dirac particles
is fully occupied\cite{footnote2}) 
against the distance from the edge $x$ measured by the magnetic length $l_B$ 
 for various values of the magnetic field with 
the energy window fixed at $-0.05 < E < 0.05$.  
In this scaled plot 
the profile of the zigzag edge states for various values of the 
magnetic fields 
fall upon common lines, where the accumulation of the charge on 
$\bullet$ sublattice as well as the depletion on $\circ$ sublattice 
are seen to occur, respectively, on the magnetic length scale toward the edge.
We have seen in Fig.\ref{phi5} that 
the flat dispersion for $B \ne 0$ on a zigzag edge 
only exists over 1/3 of the $k_2$ Brillouin zone 
(that satisfies $|1-(-1)^q e^{i q k_2}| \le 1$). 
We indeed observe that 
the accumulated $\bullet$ charge over the depleted $\circ$ 
charge, estimated by integrating the density over one half the sample width 
is $(1-\phi)/3$ within the numerical accuracy.  

\begin{figure}
\centerline{\epsfxsize=8cm\epsfbox{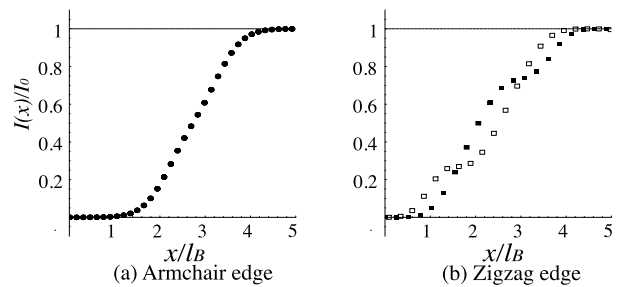}}
\caption{
Scaled plot of the charge density $I(x)$ against $x/l_B$ 
for the $n=1$ Landau level  (marked with B in Fig.~\ref{fig1}) with magnetic field 
$\phi=1/41$ for armchair(a) or zigzag(b) edges. 
} 
\label{fig3nonzero} 
\end{figure}

In order to confirm that the charge accumulation around zigzag edges 
is specific to the $n=0$ Landau level which embeds the edge mode, 
we can look at the charge density for $n=1$ Landau level.  The result 
in Fig.\ref{fig3nonzero} has $I(x)$ monotonically decreases toward the edge for both armchair and zigzag edges, although 
we can notice that 
the charge density exhibits plateaus for each 
of $\bullet$ and $\circ$ sublattices in a zigzag edge.

\begin{figure}
\vspace{-0.6cm}
\centerline{\epsfxsize=8cm\epsfbox{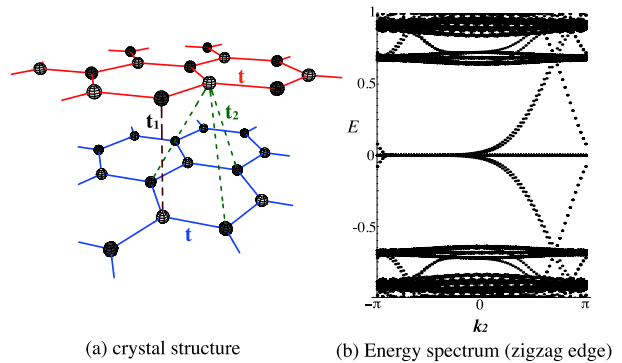}}
\caption{(Color online)
(a) A bilayer graphene with Bernal stacking 
with the top layer having a zigzag edge.  
The transfer energies considered in the 
Slonczewski-Weiss-McClure model are displayed .
(b) Energy spectrum for a bilayer graphene 
having a zigzag edge in the top layer 
with interlayer couplings $t_1=t_2= 0.1 t$ 
in a magnetic field of $\phi=1/21$.
}
\label{stack}
\end{figure}

\begin{figure}
\centerline{\epsfxsize=7cm\epsfbox{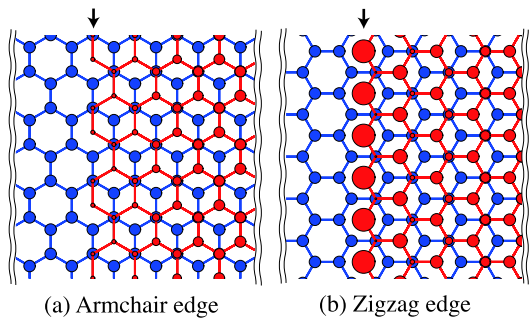}}
\caption{(Color online)
Local charge density ($\propto$ area of each circle) 
for a bilayer graphene with $t_1=t_2= 0.1 t$ 
for armchair(a) or zigzag(b) edges (indicated by arrows) 
in a magnetic field $\phi=1/21$ 
for the energy window $-0.05<E<0.05$.  
Red (blue) circles represent top (bottom) layer.
} 
\label{fig2d} 
\end{figure}

{\it Double Layer ---} 
We finally examine the bilayer graphene (Fig.~\ref{stack}(a)), 
which is interesting in its own right as many papers have pointed out, 
but the system is practically interesting as well, since, experimentally, 
the STM imaging may be easier for the edge of the top 
layer residing on a wider bottom layer.   
We consider the bilayer graphene with the AB (Bernal) stacking
in the standard 
Slonczewski-Weiss-McClure model\cite{bilayermodel1,bilayermodel2}, where 
there are two types of interlayer transfers, $t_1$ and 
$t_2$. 
For simplicity we have 
taken $t_1=t_2=0.1t$, which are roughly 
the estimated values\cite{band1,band2,band3}.  
As in the single layer, the periodic boundary condition is applied in ${\mathbf e}_2$-direction, while a wider width ($L_1=7q$) is taken for the 
bottom layer (with $L_1=5q$ for the top layer).
Despite the interlayer coupling, the Landau-quantized energy spectrum 
(Fig. ~\ref{stack}(b))
is similar to those for the $n=0$ Landau level 
on this energy scale. 
Figure~\ref{fig2d} displays the charge density for armchair and zigzag edges 
with an energy window that covers the bilayer $n=0$ Landau level. 
We can see that 
the edge states in the top layer are similar to those in the single layer, 
namely, the charge density is accumulated toward the zigzag edge on 
one sublattice.  
So we can predict that a bright edge should be observed when an 
STM study is done for a zigzag edge of the top layer in a bilayer graphene. 

To summarize, we have shown that the charge density in 
graphene in strong magnetic fields should be 
totally unlike ordinary QHE systems, where the charge is 
accumulated toward zigzag edges.   
We wish to thank Hiroshi 
Fukuyama and Tomohiro Matsui 
for illuminating discussions, and for pointing out that 
bilayer graphene may be suitable for STM imaging.  
This work has been supported in part by Grants-in-Aid for Scientific Research, 
No.20340098, 20654034 from JSPS and 
No. 220029004 on Priority Areas from MEXT for MA, 
No.20340098, 20654034 from JSPS and 
No. 220029004, 20046002 on Priority Areas from MEXT for YH, 
No.20340098 from JSPS for HA.


\begin{thebibliography}{99}
\bibitem{footnote1} 
We assume that when the left edge is armchair (zigzag), the right edge is armchair (bearded).  Although the structure of the opposite edge does not 
affect the discussion here, this convention makes the edge spectrum simpler.
\bibitem{footnote2} 
There are $2q$ magnetic Bloch bands, where each Dirac Landau band is counted 
as two contributed by K and K', so that each Dirac Landau band has 
a charge density of $1/q$.
\bibitem{footnotewidth} 
The charge density is insensitive to 
the window width as long as the $n=0$ Landau level is covered.
\end{thebibliography}
\end{document}